# Additive-hydride interaction in $MgH_2$ / $V_2O_5$ hydrogen storage composite system and its effect on electrochemical Li conversion reaction in a Li ion battery

**D. Pukazhselvan* [1,2], Ihsan Çaha[3], Francisco J.A. Loureiro[1,2], Francis Leonard Deepak[3], Catarina de Lemos[1,2], Aliaksandr L. Shaula[1,2], Sergey M. Mikhalev[1,2], Duncan Paul Fagg[1,2]**

[1]Department of Mechanical Engineering, TEMA - Centre for Mechanical Technology and Automation, University of Aveiro, Aveiro 3810-193, Portugal

[2]LASI - Intelligent Systems Associate Laboratory, Guimarães 4800-058, Portugal

[3]Nanostructured Materials Group, International Iberian Nanotechnology Laboratory (INL), Avenida Mestre Jose Veiga, Braga 4715-330, Portugal

**Abstract:**

This study explores the Li conversion characteristics of a hydrogen storage material $V_2O_5$ added $MgH_2$. Initially, to throw light on the chemical interaction between $V_2O_5$ and $MgH_2$, a detailed X ray diffraction and X ray photoelectron spectroscopy characterization study was conducted on the ball milled composite, $xMgH_2+V_2O_5$ (x=0, 0.125, 0.25, 0.5, 2, 4, 6, 8 and 10). These studies proved the chemical interaction between $MgH_2$ and $V_2O_5$ and the formation of a Mg-V combined rock salt oxide, $Mg_xV_yO_{x+y}$. It is further proved that a 5 wt.% $Mg_xV_yO_{x+y}$ added $MgH_2$ influences the following chemical reaction, $2Li+MgH_2 \rightarrow Mg+2LiH$, with a deliverable discharge / recharge capacity of 1350 mAh/g / 155 mAh/g at 0.05C in a Li ion coin cell battery. Post-mortem analyses of the used Li ion batteries prove the adequate presence of $MgH_2$, which rules out swelling issue as the reason for limited reversibility. Detailed electrochemical impedance spectroscopy analyses through differential function of relaxation times (DFRT) reveals that electrolyte degradation is not an issue in these batteries. On the other hand, DFRT provides crucial evidence for the presence of slow charge transfer processes, which are found to be sensitive to the composition of the employed additive. Based on these insights, we conclude that optimization of electrode-electrolyte compatibility is a prime factor for achieving breakthroughs in this system.

**Introduction:**

The lightweight binary hydride $MgH_2$ is a widely recognized solid state material for heat and hydrogen storage applications [1-5]. Recent research works highlight that $MgH_2$ can also be employed for storing a large quantity of electric energy (2038 mAh/g) through a Li ion battery [6-10]. At the anode side of the Li ion battery, $MgH_2$ makes a hydride conversion reaction with Li as per equation (1),

$$MgH_2 + 2Li \leftrightarrow 2LiH + Mg \qquad (1)$$

This reaction attracts interest because of its energy advantage over the standard anode material graphite (specific capacity of natural graphite: 372 mAh/g) and there exists no issues of dendrite growth. However, poor cyclic stability is a commonly reported challenge in this system [11, 12], though a cyclic reaction (1) is thermodynamic feasible. In this context, there were also efforts to understand the common causes of limited reversibility and failure mechanism because such understanding is necessary to reach a technical breakthrough in bettering the battery performance [13].

In one of the earliest efforts, Oumellal et al. [14] observed that the product of the reaction (1) is Mg embedded LiH and the experimental capacity of the battery is 1480 mAh/g. This report garnered much attention as the battery showed lower hysteresis than the other conversion type electrodes, e.g, metal oxides [15, 16], metal fluorides [17, 18], metal sulfides [19, 20] and metal phosphides [21, 22]. In their proceeding works Oumellal et al. [23] demonstrated that the size of $MgH_2$ also makes a significant impact on the system performance. For instance, an $MgH_2$ sample with a size of ~5.5 nm exhibited improved cycle life as compared to the other $MgH_2$ samples. Brutti et al. [24] employed ball milled $MgH_2$ (5 to 15h) and found that it is possible to achieve a forward conversion of ~1500 mAh/g, however a volume expansion of 200% lead to limited reversibility. Peng et al. [25] investigated the electrochemical interaction of hydrogenated Mg and Mg anchored C thin films (220 to 270 nm) with Li and found that a forward and reverse conversion of 1412 mAh/g and 518 mAh/g, respectively, can be achieved. However, there was a consistent capacity drop due to the deactivation of Mg surfaces. In another approach, El kharbachi et al. [13] tested $MgH_2$ samples ball milled for different times, 30 minutes to 24 hours, as battery material as per reaction (1) and found that a maximum forward conversion of 1450 mAh/g can be achieved, and a 38% reverse conversion can also be achieved. However, the capacity sharply decreases over cycles and no significant reversibility could be observed after 3 cycles. This team has also studied the failure mechanism of $MgH_2$ employed Li ion battery through post-mortem approach and suggested that contact loss at the carbon–LiH–Mg interfaces duly promoted by the agglomeration of Mg may be the main reason for the failure of the battery. Xu and Mulder [26] prepared a composite mixture of 5 wt.% $TiF_3$ added $MgH_2$ and CNT in 3:1 mass ratio and employed this as an active anode material in a Li ion battery. It was found that both $MgH_2$ and $TiF_3$ involve individually in the conversion reaction of Li and it was possible to deliver a capacity of 543 mAh/g. In another approach, which is to prevent the aggregation induced failure of $MgH_2$ employed Li ion battery, Zhang et al. [27] made a ~13.8 nm $MgH_2$ anchored graphene nano assembly (effective loading 50 wt.% of $MgH_2$ on graphene) and demonstrated that quick capacity degradation can be prevented, and it is possible to restore 395 mAh/g capacity.

All the above research works explored interesting insights, but the common concern is the limited reversibility. The exact reason for the limited reversibility is not well understand, however, researchers suggest two important factors, (i) phase separation and the consequent poor electrochemical contact, (ii) loss of active material from the electrode due to high volume expansion and the subsequent loss of electrochemical contact. In the literature, though several appreciable efforts were made by researchers for providing new information within the context of "(i)" and "(ii)", analyses through postmortem approaches are very limited. Particularly, there are no systematic studies aimed at exploring the differences in the electrochemical impedance of the battery in fresh state and cycled state through a DFRT approach. Considering this, in the current study, we have developed two Li ion batteries using $MgH_2$ with two different additives ($V_2O_5$ and V dissolved MgO rock salt). In these batteries after performing the required electrochemical studies (cyclic voltammetry (CV), charge/discharge test and the electrochemical impedance (EIS) test), the anode was recovered by de-crimping the battery and a postmortem analysis was performed through scanning electron microscopy. A detailed EIS analyses employing DFRT was made, which provides interesting insights regarding the points of concern with $MgH_2$ as active anode material for Li ion batteries.

**Experimental:**

Fresh $MgH_2$ samples were synthesized in our laboratory using a simple three step method (purity: >99%), as already reported in our previous study [28]. One of the additives used in the current study $V_2O_5$ was purchased from Merck Inc (99%). The other additive V dissolved MgO rock salt (typically $Mg_xV_yO_{x+y}$) was optimized after studying the interaction between $MgH_2$ and $V_2O_5$. Additives (5 wt.%) were incorporated with $MgH_2$ through mechanical milling technique using a planetary ball milling facility, Retsch PM200. The ball miller was operated at the speed of 200 rpm, at a ball to power weight ratio of 70:1 for a milling duration of 5h. A carbon substrate with a thin acrylic adhesive layer on both sides was employed for mounting the $MgH_2$ samples. Initially one side of the carbon substrate was laid over a copper current collector and in the other side the $MgH_2$ sample was mounted homogenously (effective loading: 1.2 mg/cm$^2$ and subsequently pressed under inert atmosphere. Any operation involving powder processing or battery assembling/disassembling is always performed inside an Ar filled, oxygen/moisture restricted glove box.

Two test batteries were made, which are, for convenience, referred with the codes, "A" and "B". For batteries "A" and "B", the electrode information is summarized below.

Battery A: anode active material: 5 wt.% $V_2O_5$ added $MgH_2$, counter electrode: Li metal.

Battery B: anode active material: 5 wt.% $Mg_xV_yO_{x+y}$ added $MgH_2$, counter electrode: Li metal.

For making both batteries "A" and "B" we have used Celgard 3501 separator, 1M solution of $LiPF_6$ at EC/DMC as electrolyte. Cyclic voltammetry (CV) study was performed within the potential window of 0.01 to 1.2V with the step potential of 0.5 mV at the scan rate of 0.1 mV/s. Electrochemical impedance spectroscopy (EIS) analyses were done at the frequency range of 0.01 Hz to 100 MHz. The charge/discharge tests were performed at the rate of 0.05C by using Neware battery tester (5V/50 mA) and the CV and EIS tests were made by corrtest CS100E potentiostat / galvanostat facility. To identify the chemical processes responsible for the total impedance, we have processed the experimental spectra using a distribution function of relaxation times (DFRT) procedure [29-31]. The DFRT spectra, given by a function, $R_p \cdot G(\tau)$, was obtained by solving the Fredholm equation [32, 33],

$$Z(\omega_i) = R_{ohm} + R_p \int_0^\infty \frac{\gamma(\tau)}{1+j\omega_i\tau} d\tau = R_{ohm} + R_p \int_{-\infty}^{\infty} \frac{G(\tau)}{1+j\omega_i\tau} d \ln\tau \qquad (2)$$

$$G(\tau) = \tau \cdot \gamma(\tau) \qquad (3)$$

$$\int_{-\infty}^{\infty} G(\tau)\, d \ln\tau = 1 \qquad (4)$$

Where $\tau$ is the time constant, $\omega$ is the angular frequency, $R_{ohm}$ is the ohmic resistance, $R_p$ is the total polarization resistance given by $R_p = R_T - R_{ohm}$ ($R_T$ represents the total impedance at $\omega \rightarrow 0$), and $\gamma$ (ln $\tau$ ). For the interpretation of the DFRT, a *Gaussian* fit of each of the DFRT peaks was performed [32, 33], allowing us to calculate the time constant, $\tau$, the resistance, $R$, and the capacitance, $C$, for each response (here, $\tau = RC$). DFRTs were obtained using the Tikhonov regularization (TR) strategy [32, 34]. With this approach it was possible to clearly separate and distinguish the diffusion processes (at low frequencies) from the charge-transfer processes (at higher frequencies) [31, 35]. For more details regarding these procedures the reader is referred to our previous publications [31, 35].

For XRD measurements the samples were air-sealed inside the glove box with polyimide films. We employed a Rigaku X-ray diffractometer with a CuKα radiation source with a wavelength of 1.541 Å. The X-ray photoelectron spectroscopy (XPS) measurements were performed using a Specs XR50 non-monochromated X-ray source with the Al Kα line from an aluminum cathode operating at 300W. Photoelectrons were detected using a Specs Phoibos 150 1D-DLD. The powder sample was affixed to a silicon wafer using double-sided carbon tape. For postmortem studies, the selected battery was de-crimped inside the glove box and the anode was carefully removed. The surface morphology and chemical composition of the samples were analyzed by scanning electron microscopy (SEM) equipped with energy dispersive spectroscopy (EDS) (FEI Helios NanoLab 450S Dual-Beam – FIB with UHREM FEG-SEM). FEI Helios focused ion beam (FIB) was also used to prepare the cross-section of samples. To prevent damage to the sample surface from $Ga^+$ ions during the milling process, two layers of Pt were applied. ~200 nm by electron beam and ~2 µm by ion beam. A 30 kV ion beam was used for the milling, a 3 kV electron beam was used for SEM imaging, and a 15 kV electron beam was used for the EDS analysis.

**Results and Discussion:**

In our previous work we have investigated $CrO_3$ incorporated $MgH_2$ for the purpose of (a) hydrogen storage and (b) electrochemical energy storage (Li ion batteries) [28]. Our detailed study proved that $CrO_3$ interacts with $MgH_2$ and transforms to metallic Cr by oxidizing the corresponding quantity of Mg. Such a transformation of additive is interesting as its influence reflects on its catalytic performance both for the storage of hydrogen and lithium. Moreover, these studies elevate our understanding regarding the catalytic mechanism of the metal oxides integrated metal hydride systems, which is not a well understood subject in the literature. With this background, our current focus is on $V_2O_5$ added $MgH_2$ for both, hydrogen storage and Li conversion reactions. The first part of the work, hydrogen storage using $V_2O_5$ added $MgH_2$, was already published elsewhere and the readers interested on the in-depth background information of the $V_2O_5/MgH_2$ composite system are suggested to refer this publication [36].

**X ray diffraction study:**

For the current study we have prepared the following composite samples, $xMgH_2+V_2O_5$ (x=0, 0.125, 0.25, 0.5, 2, 4, 6, 8 and 10), by ball milling them for 15h (optimal time identified by the previous study [36]). The XRD patterns corresponding to these samples are provided in Fig.1. As can be seen, under the conditions employed in the current study, clear evidence is found in all the compositions that $MgH_2$ and $V_2O_5$ interact chemically and reduce V to lower oxidation state(s). For instance, when there exists a smaller quantity of $MgH_2$ in the composite (e.g., when x = 0.125, 0.25 and 0.5), vanadium pentoxide transforms to vanadium dioxide. On the other hand, when x = 2, 4 and 6, only a monophase MgO rock salt along with a minor quantity of ball milling impurity Fe can be observed. The other Mg rich compositions (i.e., x = 8 and 10) comply well with the information noted for x= 2, 4 and 6 except that free $MgH_2$ is also present in these compositions. It is understandable because $MgH_2$ in these composites are in considerably larger quantity than $V_2O_5$. One important point to note from these XRD patterns is that no direct diffraction evidence can be found for the presence of V containing phase(s). The absence of diffraction signature from V bearing phase(s) in these cases is possible only when the situation of V falls under any of the following two categories, (i) V containing phase exists in amorphous state, (ii) V dissolved in the MgO rock salt lattice. Detailed high resolution transmission electron microscopy investigation in our previous study clarified that no amorphous regions exist in $MgH_2/V_2O_5$ composite samples. Therefore, by correlating the situations of the current study with our recent previous report [36], and also the other metal oxides added $MgH_2$ investigated by us over time [37-42], it is clear that the oxidized Mg identified in the current case is in fact V dissolved MgO rock salt, typified by a formula unit $Mg_xV_yO_{x+y}$.

**X ray photoelectron spectroscopy study:**

After conducting XRD analysis, we proceeded with XPS investigation on three samples, $xMgH_2+V_2O_5$, where x= 0.125, 2 and 4. The XPS survey spectra of these samples are provided in Fig.2a and the high resolution binding energy scans for the V2p and O1s profiles are provided in Fig.2b, c and d ("b" for $0.125MgH_2+V_2O_5$, "c" for $2MgH_2+V_2O_5$, and "d" for $4MgH_2+V_2O_5$). The complex O1s/V2p profiles were deconvoluted and the signal contribution from the individual species were marked as given in the figure. As one can see in Fig.2b ($0.125MgH_2+V_2O_5$), with a small quantity of $MgH_2$ in the powder, it is possible to transform $V^{5+}$ to $V^{4+}$, i.e., from $V_2O_5$ to $VO_2$. It is also clear that strong reduction is possible with the presence of higher quantities of $MgH_2$ together with $V_2O_5$, as can be understood from the presence of V in $V^{3+}$ / $V^{2+}$ oxidation states in the samples $2MgH_2+V_2O_5$ and $4MgH_2+V_2O_5$ (profiles Fig.2c and d). Especially in $4MgH_2+V_2O_5$, along with $V^{3+}$ and $V^{2+}$, presence of $V^0$ can also be observed. The presence of $V^{2+}$ can be directly attributed to the V dissolved MgO rock salt phase, whereas $V^{3+}$ may also be the part of this rock salt lattice. In a previous study on Cr doped MgO, Benedetti et al. [43] revealed the ability of MgO rock salt to accommodate $Cr^{3+}$ through the creation of charge compensating Mg vacancies in the lattice. Another interesting point to note from Fig.2 is the systematic change in the O1s profiles with respect to the x value in the powder $xMgH_2+V_2O_5$, which comply well with the trend observed for V in all the three spectrums. The deconvoluted O1s profile suggests the presence of three peaks, at positions 531.8 eV, 530 eV and 529.2 eV in the case of $0.125MgH_2+V_2O_5$ (Fig.2b), at positions 531.9 eV, 530 eV and 529.2 eV in the case of $2MgH_2+V_2O_5$ (Fig.2c), at positions 531.7 eV, 529.4 eV and 527 eV in the case of $4MgH_2+V_2O_5$ (Fig.2d). A clearly observable trend by comparing the peaks in profiles "b", "c" and "d" is that the high B.E peak develops from being a small shoulder to a prominent peak as the concentration of $MgH_2$ is increased in the composite. The XRD profiles demonstrated in Fig.1 also revealed the similar trend, and the responsible phase behind this trend was identified to be $Mg_xV_yO_{x+y}$. In the literature, the reported oxygen signals for MgO and $VO_x$ are at the B.E values, respectively, 531.1 eV and 530 eV [44, 45]. Therefore, by correlating with the literature evidence and the XRD observation in the current study, it is clear that the higher B.E peaks (within 531.7-531.9 eV) are due to the formation of Mg rich $Mg_xV_yO_{x+y}$ rock salt in the composite. The peaks within 530-529.4 eV can be attributed with $VO_x$ or V rich rock salt phases whereas the minor peak observed at 527 eV may be attributed to the oxygen traces existing in highly reduced cationic environment in the composite (e.g. O traces in $V^0$ rich environment). In the literature, presence of oxygen with signals at B.E closer to 527 eV is reported for a few other systems [46, 47], where the existence of oxygen under a highly cationic environment was observed.

It is clear from the XRD and XPS investigations that $V_2O_5$ and $MgH_2$ interacts chemically and produces an in-situ rock salt oxide product $Mg_xV_yO_{x+y}$. Previously tested other oxide additives, $Nb_2O_5$ and $TiO_2$ [38-42] also provided similar information (i.e., formation of $Mg_xNb_yO_{x+y}$ / $Mg_xTi_yO_{x+y}$). These are valuable insights for exploring a generalized understanding regarding how a functional metal oxide catalytically influences $MgH_2$.

**Electrochemical investigation:**

We have constructed two Li ion batteries "A" and "B" (battery "A" uses $V_2O_5$ added $MgH_2$, whereas battery "B" uses $Mg_xV_yO_{x+y}$ added $MgH_2$), and initially made a cyclic voltammetry study for both these batteries. The CV profiles recorded within the potential window of 0.01 to 1.2 V are shown in Fig.3a and b (Fig.3a for battery "A" and 3b for battery "B"). Since our samples were mounted on carbon substrate, signals from C may also be present in the CV profiles. Therefore, as a reference for comparison, CV test was also done for two more batteries: one contains no $MgH_2$ on the carbon substrate (Fig.3c) and the other contains pure $MgH_2$ (i.e. no $V_2O_5$ / $Mg_xV_yO_{x+y}$) on the carbon substate as anode (Fig.3d). In the case of battery “A” (Fig.3a), the first lithiation step (higher potential to lower potential sweep) provide peaks at the potentials, 0.6 V, 0.54 V, 0.15 V and 0.05V. During de-lithiation step (reverse sweep: lower to higher potential) three peaks were identified at the potentials, 0.18 V, 0.62V and 0.89V. Among these seven peaks, the peak at 0.15 V (forward sweep) and the one at 0.62V (reverse sweep) can be attributed to the forward and reverse Li conversion as per reaction (1). The peak pair at 0.05 V and 0.18 V, respectively, can be attributed to the alloying and de-alloying of Mg with Li, which is also confirmed by other researchers [25]. The peaks at 0.54 V and 0.89V, respectively, can be attributed to the Li insertion and extraction in the carbon substrate, whereas an irreversible interaction identifiable by a peak 0.6V can be attributed to the formation of SEI. Similarly, in the case of battery “B”, in the first cycle the lithiation step provides four peaks at 0.05 V, 0.18 V, 0.55 V and 0.7V, whereas the de-lithiation step provide peaks at the following potentials 0.18 V, 0.63 V and 0.99V. Among these the irreversible process (formation of SEI) can be identified with the peak at 0.7 V. The peaks at 0.18V and 0.63V, respectively, are due to the forward and reverse Li conversion as per reaction (1). The peaks at 0.05V and 0.18V are due to the alloying / de-alloying of $Mg/MgH_2$ with Li, whereas the peaks at 0.55 V and 0.99V can be attributed to the lithiation / de-lithiation of carbon substate alone. In the second cycle, one key difference in both the cases of battery “A” and “B” is that the forward conversion of reaction (1) occurs at slightly higher potential. It may be due to the change in the chemical surroundings in the $MgH_2$ locality that occurred during the first cycle. After performing the CV tests, we have studied the electrochemical charge/discharge characteristics of batteries “A” and “B” for five charge/discharge cycles at the rate of 0.05C (Fig.3e for battery “A” and the Fig.3f for battery “B”). As can be seen, in the case of battery “A” a forward conversion of upto 900 mAh/g can be achieved in the first cycle, whereas ~200 mAh/g can be reversibly stored. The forward / reverse conversion diminishes in the further cycles and in 5$^{th}$ cycle the restored capacity reaches 63 mAh/g. A similar trend is observed in the case of battery “B”. In this case, a forward conversion of upto 1350 mAh/g can be achieved in the first cycle whereas the restored capacity is limited to 160 mAh/g. The forward / reverse conversion in the further cycles diminishes and reaches 60 mAh/g in the 5$^{th}$ cycle. It is understandable from these observations that degradation of battery performance with cycles is a concern with both the batteries “A” and “B” which requires a rigorous analytic study. To the best of our opinion, a thorough electrochemical impedance analysis and postmortem studies may provide useful information regarding the points of concern with these batteries.

**Postmortem investigation:**

To throw light on the phase/microstructure/chemical changes occurred on the anode active materials employed in batteries “A” and “B”, three samples were further studied by XRD and SEM/EDX techniques, which are, (i) fresh anodes, i.e. anodes before subjecting to electrochemical interactions, (ii) recovered anodes after completing the first discharge, and (iii) after completing the 5 discharge/charge cycles. The observed XRDs are provided in Fig.4. In the case of fresh anodes of batteries “A” and “B”, presence of $MgH_2$ can be identified, whereas no signals from the additive phases are observed, possibly because of the small quantity of additive phases. The $MgH_2$ peaks can also be observed after the first discharge in the recovered anodes of both the batteries, “A” and “B”, however, no diffraction peaks from any products of interaction could be observed. It may be due to any of the following reasons, (i) detachment of the product of interaction from the electrode surface, or (ii), the existence of the conversion products in amorphous state. The XRDs of the recovered anodes after 5 cycles in both the cases also show the presence of $MgH_2$. From this it is understandable that the poor capacity and cycle stability issue is not because of the absence or chemical contamination of the active material $MgH_2$. The reason may be the blockage of active surfaces and the consequent electrochemical impedance. Therefore, a detailed further analysis is made through EIS technique and the results are discussed in the following sections (Fig.4-8).

The SEM images bringing out the surface morphology and cross-section for the anode used in battery "A" in fresh state (images a, b, c, d), after discharge cycle 1 (images e, f, g, h) and after completing 5 cycles (images i, j, k, l), are provided in Fig.4. The EDX elemental chemical distribution scan result obtained from the cross-section for each case are also provided in the figure (images d1, d2, d3 and d4 are the EDX scan correspond to the fresh electrode for the area provided in image d. In these, d1 is the scanned profile for Mg, V, Pt and O together, d2 for Pt, d3 for Mg and d4 for O. The images h1, h2, h3 and h4 are the EDX scan for the cycle-1 discharge tested electrode as represented in image h. In these h1 is scan for Mg, V, Pt and O together, h2 for Pt, h3 for Mg and h4 for O. Similarly, images l1, l2, l3 and l4 are the EDX scan correspond to the 5 cycles tested electrode as represented in image l. In these, l1 is the scan for Mg, V, Pt and O together, l2 for Pt, l3 for Mg and l4 for O). For the case of battery "B" similar studies were made and in Fig.5 we have provided the surface morphology, cross section image and mapping results with the similar series as given for battery "A" in Fig.4. As one can see in both the cases of batteries "A" and "B", the fresh anode shows the presence of finely distributed $MgH_2$ particles in the surface. Moreover, in both the cases $MgH_2$ particles can be identified within the depth of roughly 2 to over 5 μm (cross section images). The EDX chemical scan suggest that a large quantity of $MgH_2$ exists at the surface, hence it is clear that in both the cases of "A" and "B" $MgH_2$ is adequately available for the interaction with Li. As far as the first discharge completed batteries "A" and "B" are concerned, the surface morphology noted for the recovered anode is quite different from that observed for the fresh anode. In both the cases surface coverage of $MgH_2$ by a thick secondary layer can be noticed upon comparing images e and f. The cross-section images and the corresponding EDX images reveal the buried presence of $MgH_2$ particles (see in Fig.4 for battery "A" and Fig.5 for battery "B"). In both these cases, the $MgH_2$ particles look more enclosed and packed than when no cycle test was performed, possibly because of the penetration of electrolyte between the gaps. The SEM/EDX images of the anodes recovered after 5 cycles (Fig.4 for battery "A" and Fig.5 for battery "B"), show that the surface and bulk contain heavily agglomerated blocks of particles. The corresponding EDX chemical scan confirms that $Mg/MgH_2$ are adequately available in both the cases of "A" and "B" after 5 cycles. It suggests that physical damage (for e.g., detachment of active material through cracks, swelling so on) is not the reason for the inactivity of $Mg/MgH_2$ particles in both the cases of "A" and "B".

**Electrochemical impedance spectroscopy study through DFRT:**

To throw more light on the factors behind the performance degradation of batteries "A" and "B", we have further studied them through EIS technique. The EIS spectra were obtained under the following situations for both the batteries "A" and "B", i.e., (i) in fresh state (before subjecting to any charge/discharge operation), (ii) after completing 5 discharge/charge cycles. For comparison, EIS was also obtained for a reference battery that uses only carbon substrate as anode (i.e., no $MgH_2$ in the anode side). The obtained spectrum for the fresh reference battery, fresh battery "A" and fresh battery "B" are provided in Fig.7. With a rough overview of the raw spectra, it is understandable that there exist signs for the slow processes in all these batteries, typically observable at lower frequencies (for clarity, a magnified version of the spectra with highlights to the transition resistance is given in the inset). Nonetheless, in all the three cases, since the signals from the individual processes are overlapped, we have deconvoluted the observed complex impedance spectra. The spectrum was reconstructed through a subtraction procedure, and then separated to individual components by introducing

DFRT. For more information regarding the theoretical background of DFRT and the subtraction procedure, we recommend readers the following references [32, 33].

In Fig.7 the raw EIS spectra (left) and the corresponding DFRT (right) are provided in the following order, (a) the fresh reference battery, (b) the fresh battery "A", and (c) the fresh battery "B". The individual resistive components of the total polarization resistance (Rp) were separated and the observed four resistance contributions (R1, R2, R3, and R4) along with the corresponding time constant and capacitance values are summarized in Table-1. Four potential contributions to total polarization resistance (Rp), typically indicated by P1, P2, P3, and P4 are due to charge-transfer interactions occurring at the surface of the electroactive particles, as denoted by the C values lying between $10^{-6}$ to $10^{-4}$ F $cm^{-1}$ [29, 31]. These charge transfer processes may include the opposition to the charge released at Li, the resistance encountered when charge moves between the electrolyte and the electrode, and the resistance at the interfaces (such as in SEI, interface between the current collector and the active electrode materials). As can be understood by comparing with the reference battery, in each case of "A" and "B", all the four resistance contributions (R1, R2, R3 and R4) are higher than in reference battery. In contrast, the ohmic resistance in batteries "A" and "B" remains the same as the reference battery, hence it is clear that electrolyte pose no performance degradation issues in these batteries. From these, it is undeniable that the resistive character increases with contribution from each of P1, P2, P3 and P4 in both batteries "A" and "B", due to $MgH_2$. Moreover, it is also notable through the DFRTs provided in "b2" and "c2" that the additive has a considerable influence on the relaxation time in each resistive process. In Fig.8, profile "(a)", "(b)" and "(c)", respectively, brings out the observed EIS spectra and the corresponding DFRTs of the 5 cycles tested reference battery, battery "A" and "B". The observed ohmic resistance ($\boldsymbol{R}_{\mathbf{ohm}}$), the net polarization resistance ($\boldsymbol{R}_{\mathbf{p}}$) and the individual resistance contributions are summarized in Table – 2. The observed ohmic resistance ($R_{ohm}$) in all the cases suggest that no significant issues occurred with electrolytes in all the batteries. However, the high $\boldsymbol{R}_{\mathbf{p}}$ values, especially in the cases of batteries "A" and "B" suggest that the active material ($MgH_2$) faces higher charge transfer restrictions during the course of charge/discharge operations. The corresponding DFRTs, as provided in b2 and c2 suggest that, despite the comparatively less resistance in the fresh state of "A", after cycles it shows higher resistance and signs of slow processes than in "B". These observations are in compliance with the charge / discharge characteristics demonstrated in Fig.3.

**Discussions:**

It is clear from the XRD and XPS sections of the current study that the additive $V_2O_5$ interacts with $MgH_2$ and produces a rock salt product, $Mg_xV_yO_{x+y}$. The results of our comprehensive hydrogen storage experiments published elsewhere [36] suggest that the in-situ formation of the $Mg_xV_yO_{x+y}$ rock salt phase makes a positive impact on the low-temperature dehydrogenation behaviour of $MgH_2$. We tested the similar possibility for the Li conversion reaction in this study through a comparison of a 5 wt.% $V_2O_5$ and $Mg_xV_yO_{x+y}$ added $MgH_2$ as anode active materials in a lithium-ion battery. Improved results are observed as compared to the case of pure $MgH_2$ discussed in our previous study [28], but cyclic stability remains a concern in the current study. When we add a small quantity of $V_2O_5$ with $MgH_2$, since there exists a chemical reaction leading to the formation of $Mg_xV_yO_{x+y}$, in fact the $V_2O_5$ added $MgH_2$ itself may contains $Mg_xV_yO_{x+y}$ active species. In this context the chemical nature of the active species existing in battery "A" and battery "B" is the same. However, there can be differences in the x and y values in both cases, which would reflect in their catalytic performance, as we have already demonstrated such catalytic differences in one of our earlier studies [39]. In the current study, the DFRT provided interesting clues that the low frequency, charge transfer processes are distinctly influenced by the active species present in battery “A” and battery “B”. Another interesting note from the current study is that the electrolyte remains unaffected, which means the detrimental effect is probably due to the obstruction of active sites in the $MgH_2$ surface or the surface coverage by electrochemically poorly active surface barriers. To address these, in our opinion, the future studies should focus on the modified electrode fabrication approach, with particular attention paid to the optimization of electrode-electrolyte compatibility.

## Conclusions:

The following conclusions can be derived from the current study,

1. Incorporation of the additive $V_2O_5$ with $MgH_2$ results in a chemical interaction that leads to the formation of a rock salt phase $Mg_xV_yO_{x+y}$.
2. As per CV observations, a 5 wt.% of $V_2O_5$ and $Mg_xV_yO_{x+y}$ added $MgH_2$ electrochemically interacts with $MgH_2$ though a two-step reaction. The potential at which the reaction occurs is closely identical in both the cases.
3. The first discharge cycle in the case of $Mg_xV_yO_{x+y}$ added $MgH_2$ containing battery is 1350 mAh/g which is higher than that observed for $V_2O_5$ added $MgH_2$ employed battery (i.e. 900 mAh/g). However, cyclic stability remains a concern in both the cases (employed charge/discharge rate 0.05C).
4. Post-mortem study though EDX technique confirms the presence of $MgH_2$ in the cycles tested sample, hence loss of $MgH_2$ due to issues like swelling is not likely to be the reason for limited reversibility.
5. As per the EIS study, the electrolyte presents no degradation issue in the case of $V_2O_5$ and $Mg_xV_yO_{x+y}$ added $MgH_2$ electrodes. However, a sharp increase in charge transfer resistance after a few charge / discharge cycles is a common issue in both the cases.

## Acknowledgement

This article was supported by the projects 2022.09319.PTDC (https://doi.org/10.54499/2022.09319.PTDC), 2022.02498.PTDC (https://doi.org/10.54499/2022.02498.PTDC), UIDB/00481/2020 (https://doi.org/10.54499/UIDB/00481/2020), and UIDP/00481/2020 (https://doi.org/10.54499/UIDP/00481/2020) from Fundação para a Ciência e a Tecnologia (FCT), and CENTRO-01-0145-FEDER-022083 from Centro Portugal Regional Operational Programme (Centro2020), under the PORTUGAL 2020 Partnership Agreement, through the European Regional Development Fund (ERDF). D. P and F. L acknowledge FCT, Portugal for the financial support with reference CEECIND/04158/2017 (https://doi.org/10.54499/CEECIND/04158/2017/CP1459/CT0029) and 2020.02797.CEECIND/CP1589/CT0030 (https://doi.org/10.54499/2020.02797.CEECIND/CP1589/CT0030), respectively. This article is also a result of the Innovation Pact “NGS - New Generation Storage” (C644936001-00000045), by “NGS” Consortium, co-financed by Next Generation EU, through the Incentive System “Agendas para a Inovação Empresarial” (“Agendas for Business Innovation”), within the Recovery and Resilience Plan (PRR). This project has also received funding from the SMART-ER project, funded by the European Union's Horizon 2020 research and innovation programme under Grant Agreement #101016888. The authors are also grateful for the financial support granted by the Recovery and Resilience Plan (PRR) and, by the Next Generation EU European Funds to Universidade de Aveiro, through the Agenda for Business Innovation “NGS - Next Generation Storage” (Project no 02/C05-i01.01/2022 with the application C644936001-00000045).

**Figure captions:**

Fig.1 XRD profiles obtained for the15h ball milled $xMgH_2+V_2O_5$ (x=0, 0.125, 0.25, 0.5, 2, 4, 6, 8 and 10) composite ($V_2O_5$ correspond to x=0).

Fig.2 XPS (a) survey spectra corresponding to $xMgH_2+V_2O_5$ (x=0.125, 2 and 4). The V2p and O1s high resolution binding energy profiles obtained for the samples, (b) $0.125MgH_2+V_2O_5$, (c) $2MgH_2+V_2O_5$ and (d) $4MgH_2+V_2O_5$.

Fig.3 Cyclic voltammetry profiles corresponding to the battery employing (a) 5 wt.% $V_2O_5$ added $MgH_2$ as anode active material (battery “A”), (b) 5 wt.% $Mg_xV_yO_{x+y}$ added $MgH_2$ as anode active material (battery “B”), (c) carbon substrate alone as anode active material and (d) pure $MgH_2$ (additive free) as anode active material. Five cycles charge/discharge profiles corresponding to (e) battery “A” and, (f) battery “B”.

Fig.4 XRD profiles corresponding to (a) the fresh anode used in battery “A”, (b) recovered anode from “A” after first discharge, (c) recovered anode from “A” after $5^{th}$ charging cycle, (d) the fresh anode used in battery “B”, (e) recovered anode from “B” after first discharge, (f) recovered anode from “B” after $5^{th}$ charging cycle. The XRD profile of Cu current collector is provided for comparison.

Fig.5 SEM / EDX profiles of anodes employed in battery “A”. Images (a), (b), (c) and (d), fresh anode (surface and cross sections). Images d1, d2, d3 and d4 correspond to EDS maps of cross section region given in “(d)” (d1 is collective distribution of Mg, V, Pt and O, d2 is Pt alone, d3 is Mg alone and d4 is O). In similar sequence images, (e), (f), (g), (h), h1, h2, h3 and h4 correspond to SEM/EDX of electrode recovered from battery “A” after first discharge. The images (i), (j), (k), (l), l1, l2, l3 and l4 correspond to SEM/EDX of electrodes recovered from battery “A” after fifth recharge cycle.

Fig.6 SEM / EDX profiles of anodes employed in battery “B”. Images (a), (b), (c) and (d), fresh anode (surface and cross sections). Images d1, d2, d3 and d4 correspond to EDS maps of cross section region given in “(d)” (d1 is collective distribution of Mg, V, Pt and O, d2 is Pt alone, d3 is Mg alone and d4 is O). In similar sequence images, (e), (f), (g), (h), h1, h2, h3 and h4 correspond to SEM/EDX of electrode recovered from battery “B” after first discharge. The images (i), (j), (k), (l), l1, l2, l3 and l4 correspond to SEM/EDX of electrodes recovered from battery “B” after fifth recharge cycle.

Figure 7 – (a1), (b1), and (c1) Complex plane impedance spectra obtained for the fleshly-made batteries (the high-magnification inset includes the subtracted polarization branch. The numbers represent the decades ($log_{10}f$) of the measuring frequency); (a2), (b2), and (c2) the corresponding DFRTs of the polarization branch.

Figure 8 – (a1), (b1), and (c1) Complex plane impedance spectra obtained for the 5 cycles batteries (the high-magnification inset includes the subtracted polarization branch. The numbers represent the decades ($log_{10}f$) of the measuring frequency); (a2), (b2), and (c2) the corresponding DFRTs of the polarization branch.

**Tables:**

Table 1 – EIS data (resistance (*R*), time constant ($\tau_0$), and capacitance (*C*)) for the freshly-made batteries.

| **Parameters** | | **Fresh batteries** | | |
|---|---|---|---|---|
| | | **No $MgH_2$ (only C substrate)** | **$MgH_2$+5%$V_2O_5$ (battery "A")** | **$MgH_2$+5%$Mg_xV_yO_{x+y}$ (battery "B")** |
| **$R_{ohm}$ (Ω cm$^2$)** | | 4.88 | 5.72 | 5.41 |
| **$R_p$ (Ω cm$^2$)** | | 838.72 | 1004.39 | 2573.76 |
| **P1** | **$R_1$ (Ω cm$^2$)** | 18.67 | 4.77 | 64.54 |
| | **$\tau_1$ (s)** | 1.34 x $10^{-4}$ | 1.55 x $10^{-5}$ | 4.41 x $10^{-4}$ |
| | **$C_1$ (F cm$^{-2}$)** | 7.18 x $10^{-6}$ | 3.25 x $10^{-6}$ | 6.84 x $10^{-6}$ |
| **P2** | **$R_2$ (Ω cm$^2$)** | 194.47 | 49.40 | 225.68 |
| | **$\tau_2$ (s)** | 3.22 x $10^{-3}$ | 2.11 x $10^{-4}$ | 3.41 x $10^{-3}$ |
| | **$C_2$ (F cm$^{-2}$)** | 1.66 x $10^{-5}$ | 4.25 x $10^{-6}$ | 1.51 x $10^{-5}$ |
| **P3** | **$R_3$ (Ω cm$^2$)** | 232.93 | 428.22 | 1564.20 |
| | **$\tau_3$ (s)** | 2.63 x $10^{-2}$ | 6.01 x $10^{-3}$ | 5.82 x $10^{-2}$ |
| | **$C_3$ (F cm$^{-2}$)** | 1.13 x $10^{-4}$ | 1.40 x $10^{-5}$ | 3.72 x $10^{-5}$ |
| **P4** | **$R_4$ (Ω cm$^2$)** | 392.65 | 521.99 | 719.35 |
| | **$\tau_4$ (s)** | 4.37 x $10^{-1}$ | 8.18 x $10^{-2}$ | 6.32 x $10^{-1}$ |
| | **$C_4$ (F cm$^{-2}$)** | 1.13 x $10^{-3}$ | 1.57 x $10^{-4}$ | 8.78 x $10^{-4}$ |

**Table 2** – EIS data (resistance (*R*), time constant ($\tau_0$), and capacitance (*C*)) for the 5 cycles tested batteries.

| **Parameters** | | **Five cycles tested batteries** | | |
|---|---|---|---|---|
| | | **No $MgH_2$ (only C substrate)** | **$MgH_2$+5%$V_2O_5$ (battery "A")** | **$MgH_2$+5%$Mg_xV_yO_{x+y}$ (battery "B")** |
| **$R_{ohm}$ ($\Omega$ cm$^2$)** | | 5.17 | 6.70 | 8.07 |
| **$R_p$ ($\Omega$ cm$^2$)** | | 1619.13 | 9038.18 | 5765.55 |
| **P1** | **$R_1$ ($\Omega$ cm$^2$)** | 5.63 | 124.20 | 33.87 |
| | **$\tau_1$ (s)** | $2.37 \times 10^{-5}$ | $4.05 \times 10^{-4}$ | $4.96 \times 10^{-5}$ |
| | **$C_1$ (F cm$^{-2}$)** | $4.22 \times 10^{-6}$ | $3.26 \times 10^{-6}$ | $1.47 \times 10^{-6}$ |
| **P2** | **$R_2$ ($\Omega$ cm$^2$)** | 135.87 | 680.58 | 210.87 |
| | **$\tau_2$ (s)** | $4.29 \times 10^{-4}$ | $5.68 \times 10^{-3}$ | $6.03 \times 10^{-4}$ |
| | **$C_2$ (F cm$^{-2}$)** | $3.16 \times 10^{-6}$ | $8.34 \times 10^{-6}$ | $4.03 \times 10^{-6}$ |
| **P3** | **$R_3$ ($\Omega$ cm$^2$)** | 885.54 | 1226.15 | 1988.30 |
| | **$\tau_3$ (s)** | $1.00 \times 10^{-2}$ | $5.20 \times 10^{-2}$ | $1.77 \times 10^{-2}$ |
| | **$C_3$ (F cm$^{-2}$)** | $1.13 \times 10^{-5}$ | $4.24 \times 10^{-5}$ | $1.25 \times 10^{-5}$ |
| **P4** | **$R_4$ ($\Omega$ cm$^2$)** | 592.09 | 7007.26 | 5885.00 |
| | **$\tau_4$ (s)** | $6.16 \times 10^{-2}$ | $9.14 \times 10^{-1}$ | $1.86 \times 10^{-1}$ |
| | **$C_4$ (F cm$^{-2}$)** | $1.04 \times 10^{-4}$ | $1.30 \times 10^{-4}$ | $4.47 \times 10^{-5}$ |

# Figures:

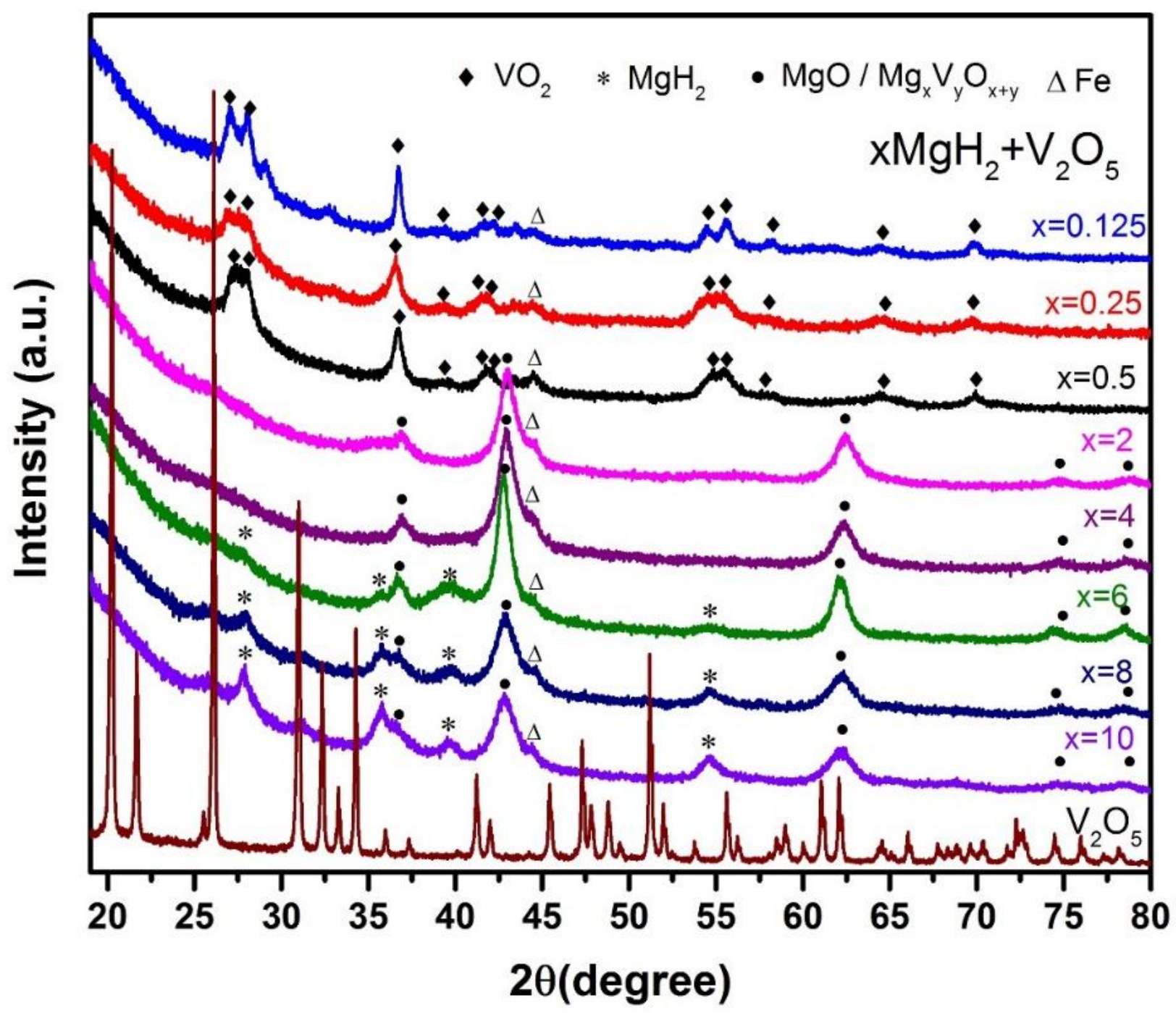

Fig.1 XRD profiles obtained for the15h ball milled $xMgH_2+V_2O_5$ (x=0, 0.125, 0.25, 0.5, 2, 4, 6, 8 and 10) composite ($V_2O_5$ correspond to x=0).

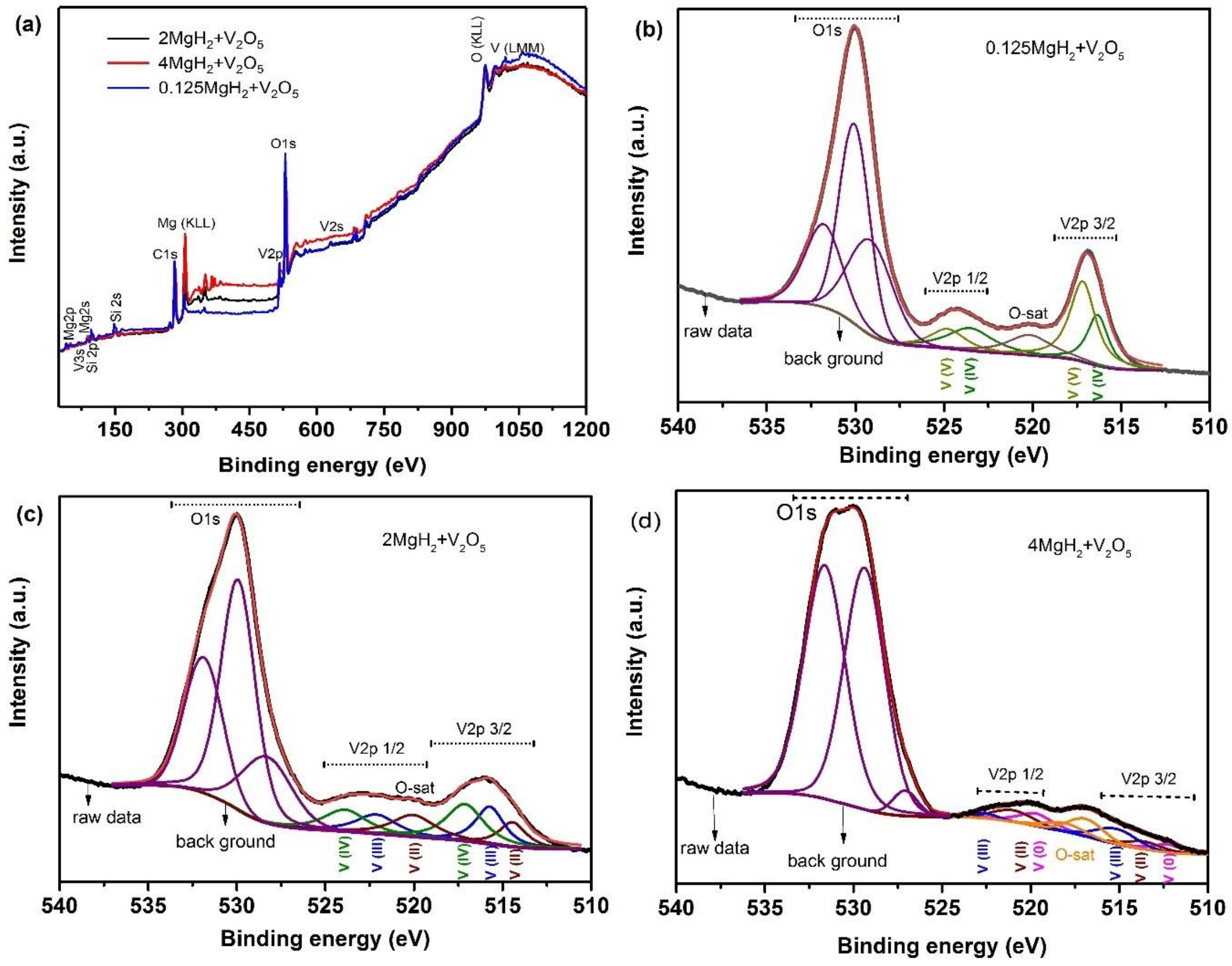

Fig.2 XPS (a) survey spectra corresponding to $xMgH_2+V_2O_5$ (x=0.125, 2 and 4). The V2p and O1s high resolution binding energy profiles obtained for the samples, (b) $0.125MgH_2+V_2O_5$, (c) $2MgH_2+V_2O_5$ and (d) $4MgH_2+V_2O_5$.

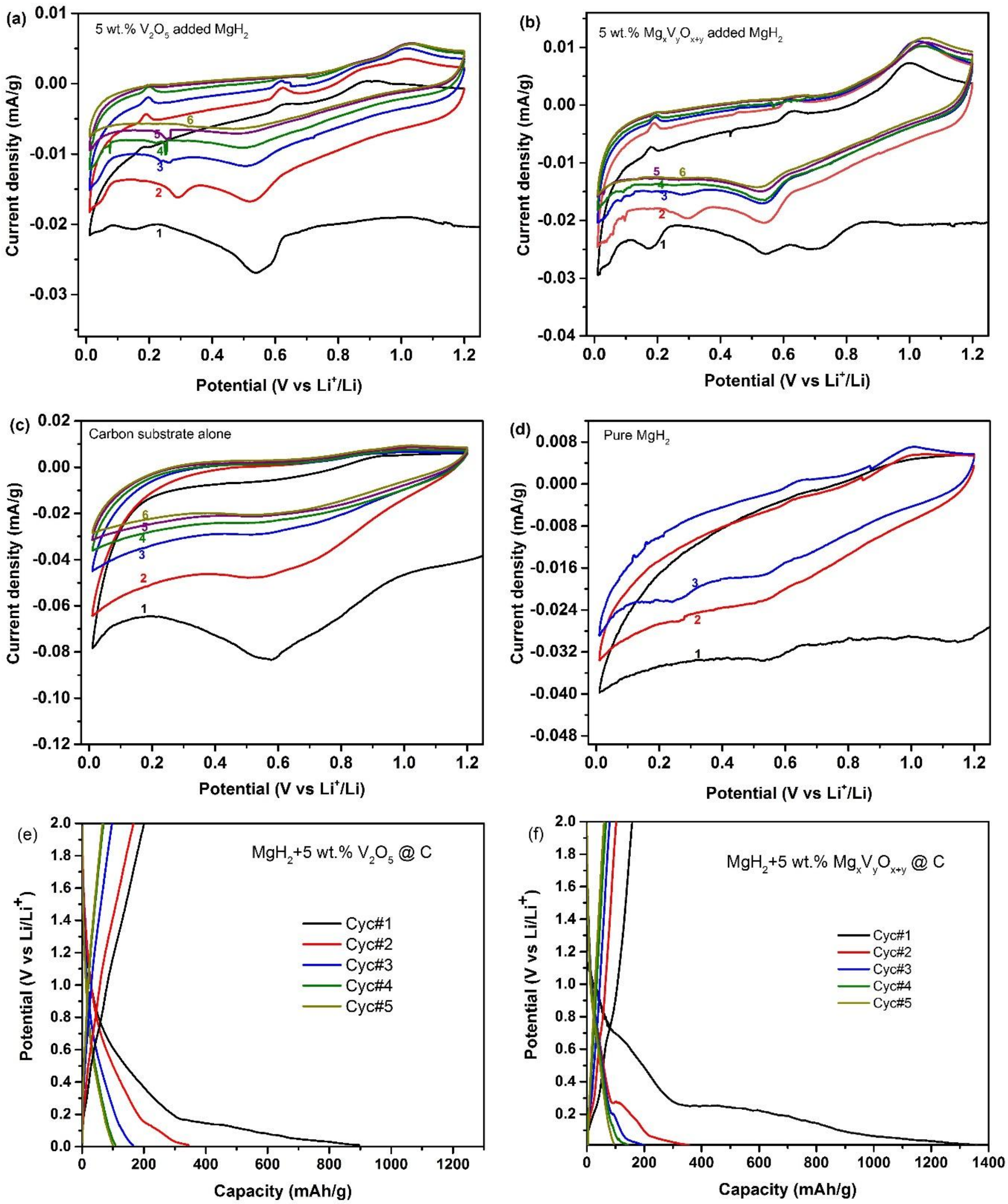

Fig.3 Cyclic voltammetry profiles corresponding to the battery employing (a) 5 wt.% $V_2O_5$ added $MgH_2$ as anode active material (battery "A"), (b) 5 wt.% $Mg_xV_yO_{x+y}$ added $MgH_2$ as anode active material (battery "B"), (c) carbon substrate alone as anode active material and (d) pure $MgH_2$ (additive free) as anode active material. Five cycles charge/discharge profiles corresponding to (e) battery "A" and, (f) battery "B".

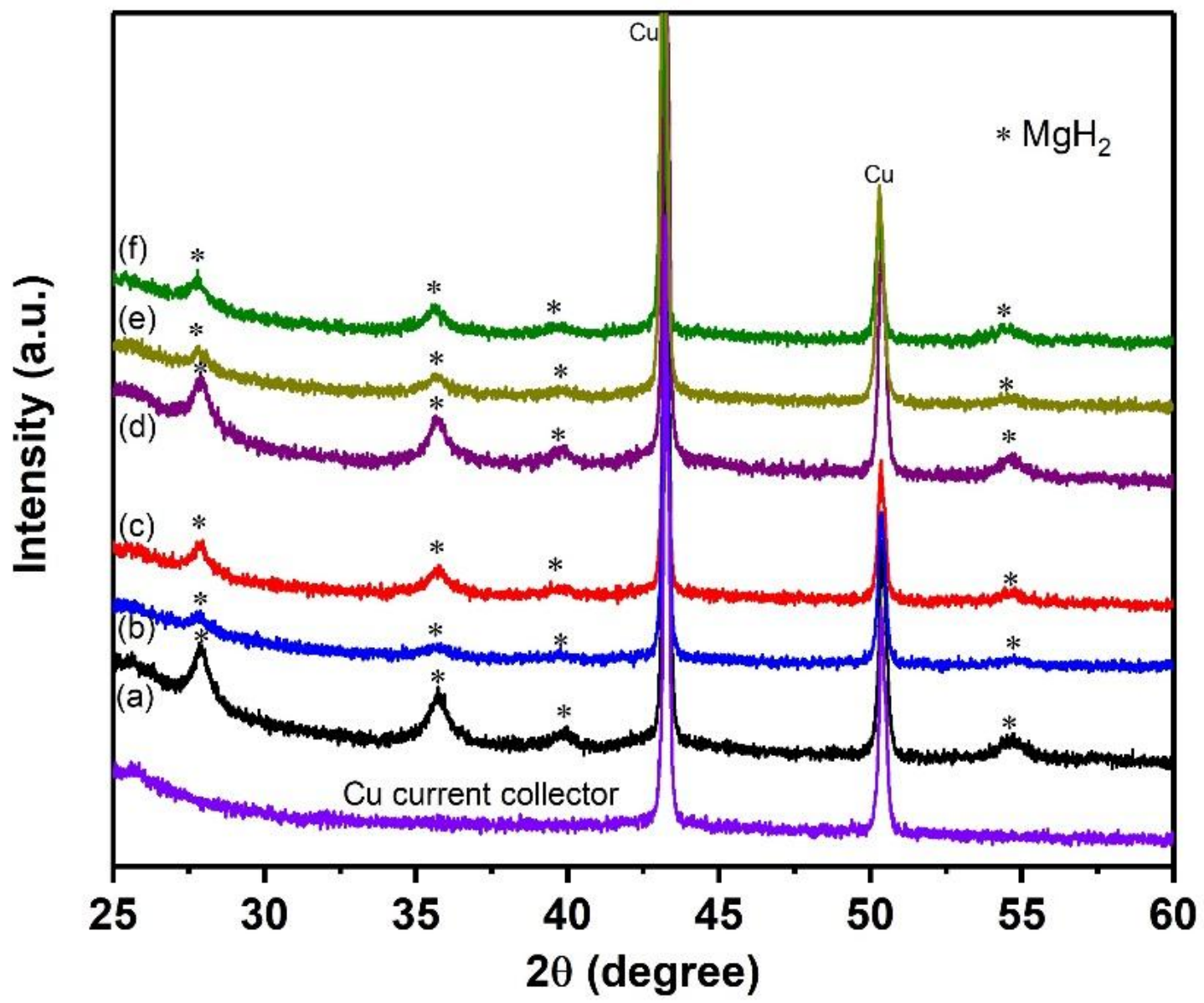

Fig.4 XRD profiles corresponding to (a) the fresh anode used in battery “A”, (b) recovered anode from “A” after first discharge, (c) recovered anode from “A” after $5^{th}$ charging cycle, (d) the fresh anode used in battery “B”, (e) recovered anode from “B” after first discharge, (f) recovered anode from “B” after $5^{th}$ charging cycle. The XRD profile of Cu current collector is provided for comparison.

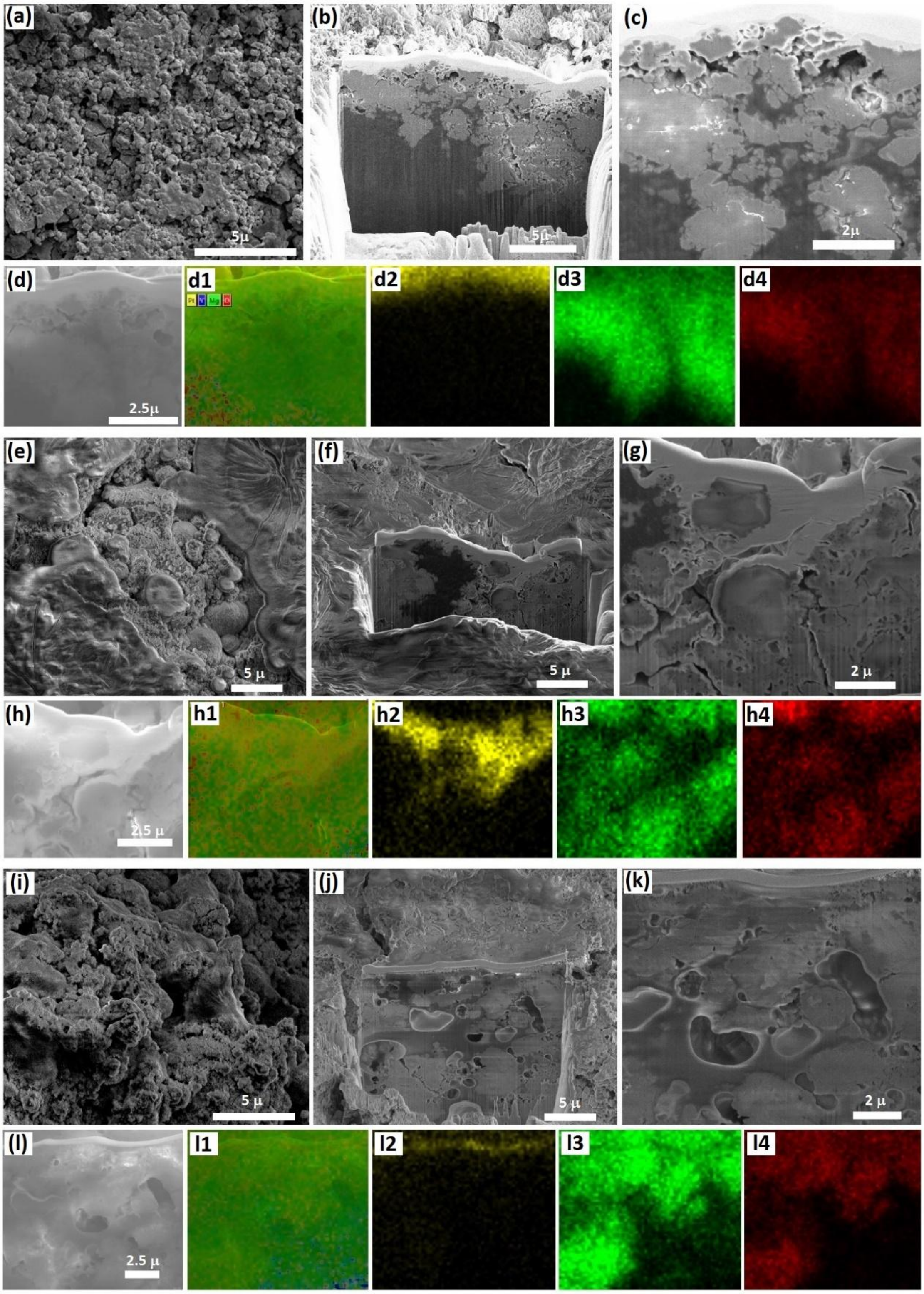

Fig.5 SEM / EDX profiles of anodes employed in battery "A". Images (a), (b), (c) and (d), fresh anode (surface and cross sections). Images d1, d2, d3 and d4 correspond to EDS maps of cross section region given in "(d)" (d1 is collective distribution of Mg, V, Pt and O, d2 is Pt alone, d3 is Mg alone and d4 is O). In similar sequence images, (e), (f), (g), (h), h1, h2, h3 and h4 correspond to SEM/EDX of electrode recovered from battery "A" after first discharge. The images (i), (j), (k), (l), l1, l2, l3 and l4 correspond to SEM/EDX of electrodes recovered from battery "A" after fifth recharge cycle.

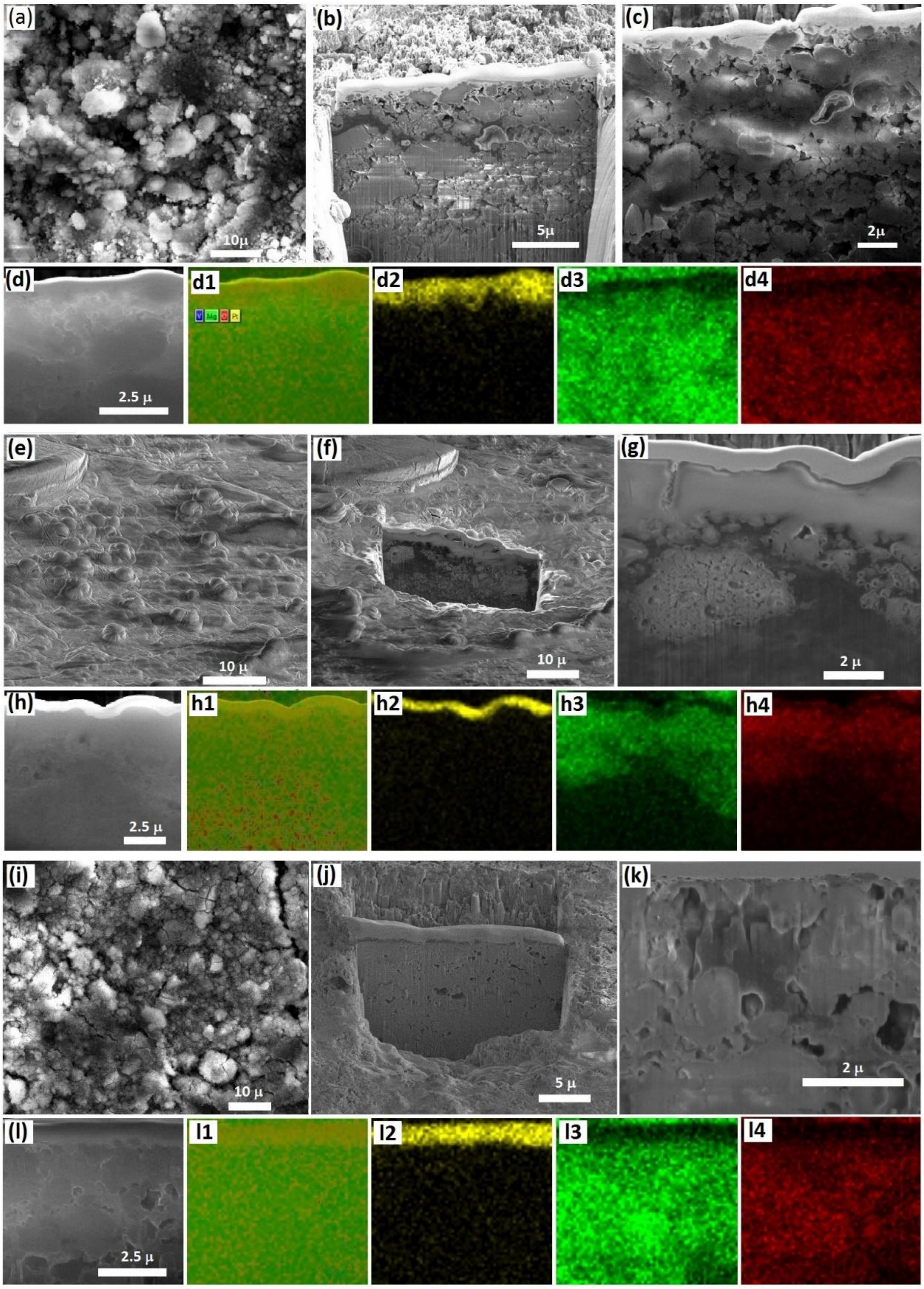

Fig.6 SEM / EDX profiles of anodes employed in battery "B". Images (a), (b), (c) and (d), fresh anode (surface and cross sections). Images d1, d2, d3 and d4 correspond to EDS maps of cross section region given in "(d)" (d1 is collective distribution of Mg, V, Pt and O, d2 is Pt alone, d3 is Mg alone and d4 is O). In similar sequence images, (e), (f), (g), (h), h1, h2, h3 and h4 correspond to SEM/EDX of electrode recovered from battery "B" after first discharge. The images (i), (j), (k), (l), l1, l2, l3 and l4 correspond to SEM/EDX of electrodes recovered from battery "B" after fifth recharge cycle.

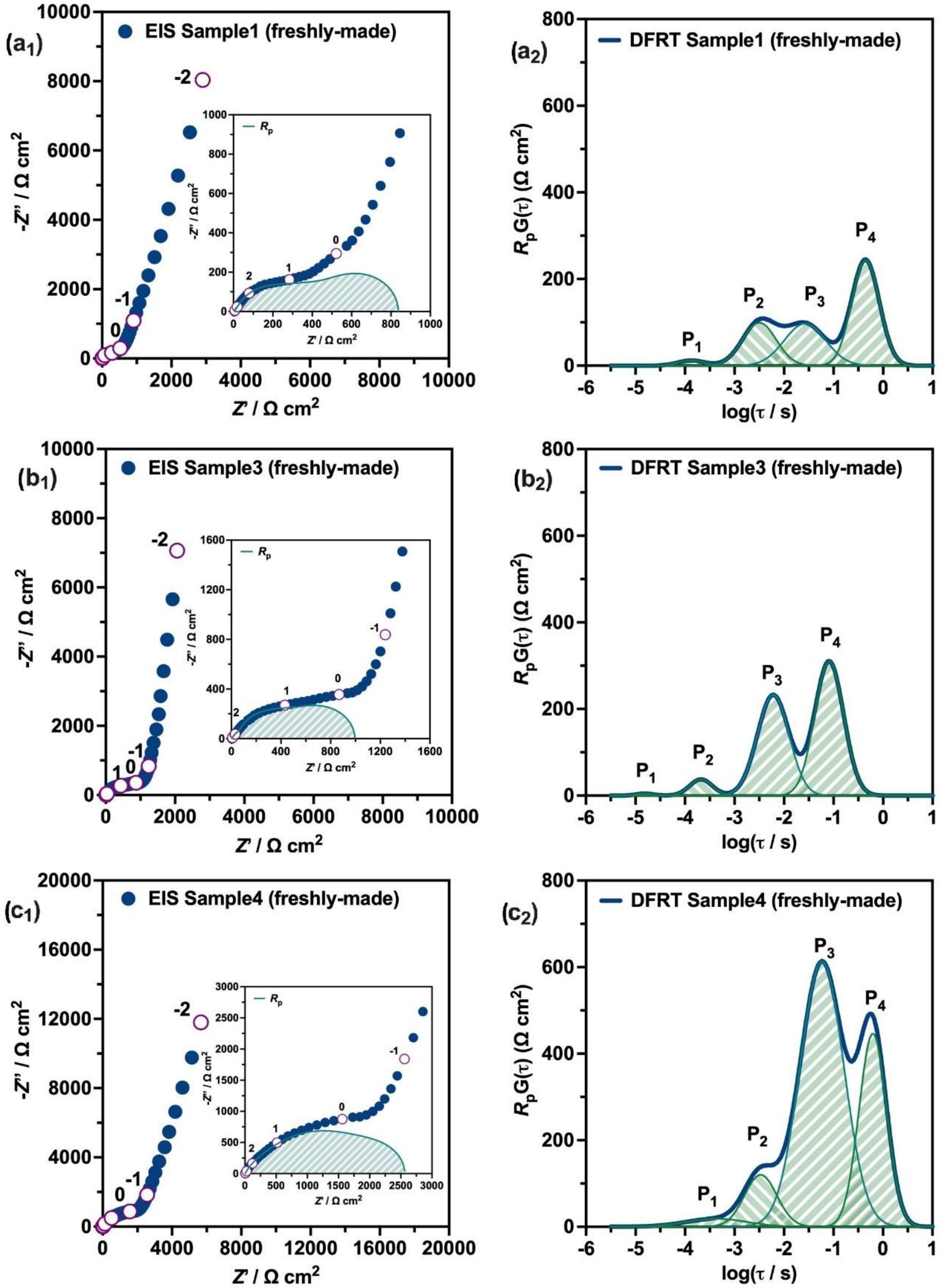

Figure 7 – (a1), (b1), and (c1) Complex plane impedance spectra obtained for the fleshly-made batteries (the high-magnification inset includes the subtracted polarization branch. The numbers represent the decades ($\log_{10}f$) of the measuring frequency); (a2), (b2), and (c2) the corresponding DFRTs of the polarization branch.

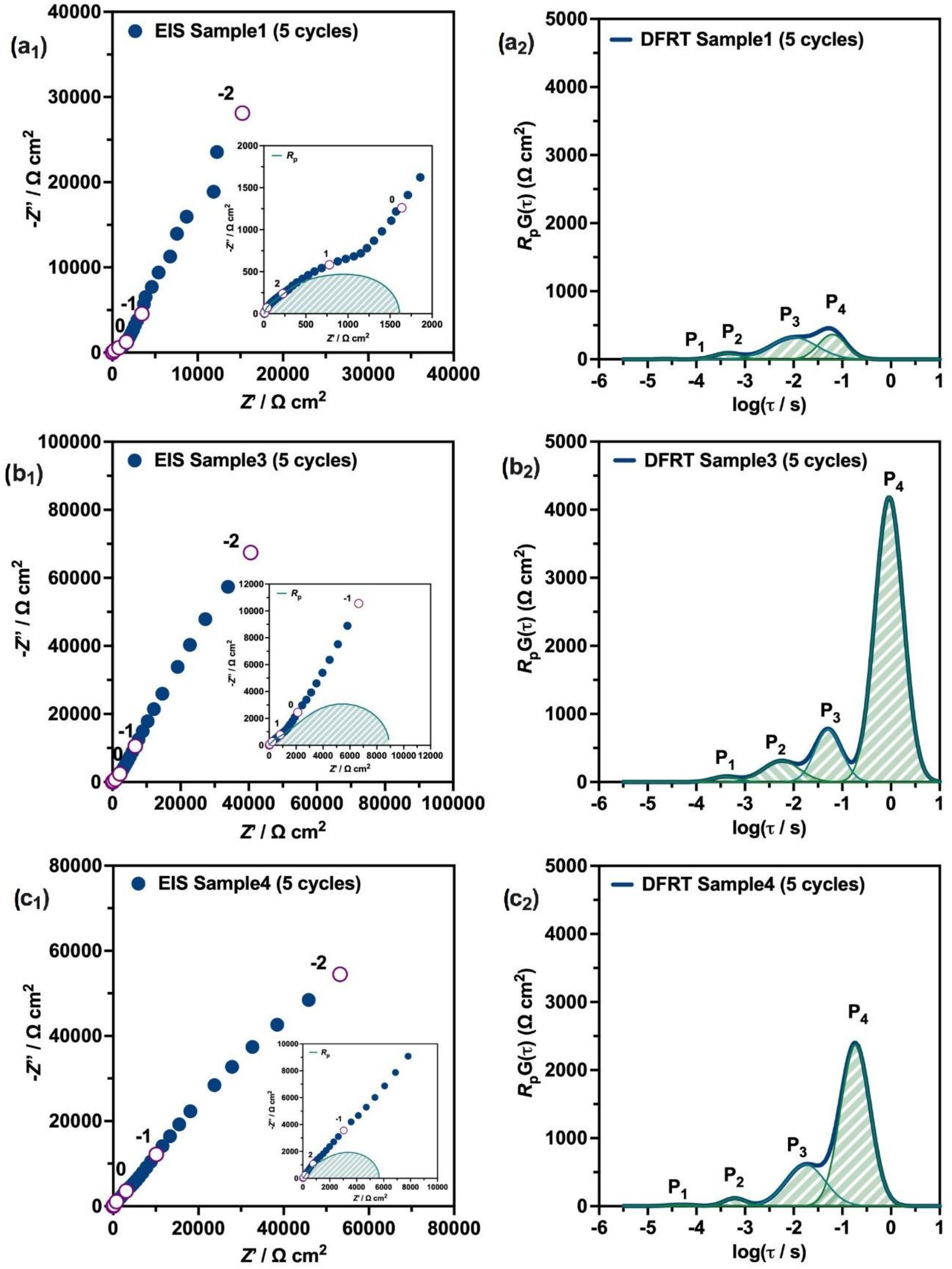

Figure 8 – (a1), (b1), and (c1) Complex plane impedance spectra obtained for the 5 cycles batteries (the high-magnification inset includes the subtracted polarization branch. The numbers represent the decades ($\log_{10}$f) of the measuring frequency); (a2), (b2), and (c2) the corresponding DFRTs of the polarization branch.